RESEARCH ARTICLE

# Screening method for early dementia using sound objects as voice biomarkers

Adam Pluta[1]  |  Zbigniew Pioch[1]  |  Jędrzej Kardach[1]  |  Piotr Zioło[3]  |  Tomasz Kręcicki[2]  |  Elżbieta Trypka[1,3]

[1] Vivid Mind

[2] Clinical Department of Otolaryngology, Head and Neck Surgery, Wroclaw Mecial University

[3] Department of Psychiatry, Wroclaw Medical University

[4] Faculty of Mathematics and Computer Science, Adam Mickiewicz University

## Abstract

**Introduction:** We present a screening method for early dementia using features based on sound objects as voice biomarkers.

**Methods:** The final dataset used for machine learning models consisted of 266 observations, with a distribution of 186 healthy individuals, 46 diagnosed with Alzheimer's, and 34 with MCI. This method is based on six-second recordings of the sustained vowel /a/ spoken by the subject. The main original contribution of this work is the use of carefully crafted features based on sound objects. This approach allows one to first represent the sound spectrum in a more accurate way than the standard spectrum, and then build interpretable features containing relevant information about subjects' control over their voice.

**Results:** ROC AUC obtained in this work for distinguishing healthy subjects from those with MCI was 0.85, while accuracy was 0.76. For distinguishing between healthy subjects and those with either MCI or Alzheimer's the results were 0.84, 0.77, respectively.

**Conclusion:** The use of features based on sound objects enables screening for early dementia even on very short recordings of language-independent voice samples.

# 1 | INTRODUCTION

Good healthcare, a higher quality of life, and an extended lifespan contribute to a new challenge – the demographic problem of aging societies. As a consequence, there is a higher frequency of dementia, especially Alzheimer's disease (AD). Currently, nearly 46 million people suffer from some form of dementia, and it is projected that by 2050, there will be 131.5 million worldwide [Pri15].

The progression of the disease leads to a reduction in occupational, social, and family activities. Dementia diseases involve caregivers and require significant financial resources at both the state and family levels. Therefore, most research efforts are currently focused on searching for not only new treatments but also new objective screening diagnostic tests for early detection of cognitive function disorders. This will enable a more personalized treatment plan. For many years, there has been a search for connections between changes in sensory organs and the onset of dementia. One of the earliest changes observed in the initial stages of dementia is voice changes. These voice changes can manifest in various ways and they can have a significant impact on communication. Some common voice changes associated with Alzheimer's disease are connected to slurred or slow speech resulting from the affected motor control and coordination required for clear speech.

As a result, individuals with Alzheimer's may speak more slowly and may slur their words. Alzheimer's disease can also affect the emotional tone and prosody (the rhythm and intonation) of speech. Some individuals may speak in a flat or monotonous tone, while others may exhibit emotional outbursts or inappropriate emotional responses during conversation [Yan22]. Alzheimer's can affect the brain's ability to process and produce speech. This impact can be observed in the simplification of spoken language, where complex sounds become more difficult to articulate, making the analysis of single vowels relevant. When a person utters single vowels, subtle changes in voice quality, pitch, loudness, and duration may occur. In Alzheimer's, the control over muscles involved in speech production can be impaired, affecting how vowels are pronounced [Haj23]. The mechanisms underlying changes in Alzheimer's disease are complex and not fully understood. These changes are part of a broader spectrum of language and communication difficulties associated with the condition. Understanding the underlying mechanisms can help in the diagnostic process, especially in the preclinical stages of disease. Language and communication involve a complex network of brain regions that work together to produce, understand, and process spoken and written language. The key brain regions involved in language and communication:

- Broca's Area located in the left frontal lobe is responsible for speech production and language processing. Damage to this area can result in difficulties in forming grammatically correct sentences and speech production.
- Wernicke's Area, situated in the left temporal lobe, is primarily responsible for language comprehension and understanding spoken and written language. Damage to this area can lead to difficulties in understanding language and producing coherent speech.
- Arcuate fasciculus connects Broca's and Wernicke's areas, facilitating communication between the language production and comprehension regions. It plays a crucial role in the integration of language functions.
- Found in the temporal lobe, the primary auditory cortex is responsible for processing auditory information, including speech sounds. It plays a vital role in perceiving and distinguishing speech sounds.
- The supramarginal gyrus is situated in the parietal lobe and is involved in phonological processing, which includes recognizing and manipulating speech sounds.
- The cerebellum is increasingly recognized for its role in language processing, including fine-tuning speech-motor control.
- Located in the parietal lobe, the angular gyrus is involved in the processing of written language, including reading and writing. It helps to link visual information with language comprehension.
- The next structure involved in the speech pro-

cess is the inferior parietal lobule. This region is responsible for various aspects of language processing, including semantic processing, understanding word meanings, and interpreting complex sentence structures.

- The prefrontal cortex plays a role in higher-level language functions, such as language planning, executive control, and decision-making related to language use.
- The amygdala is involved in the emotional aspects of language processing. It helps to attach emotional significance to words and language.
- And finally, the hippocampus plays a role in forming new memories related to language, including vocabulary and contextual information. [Hic07] [Pri12] [Hag14] [Cha10] [Fri13]

These brain regions tend to support various aspects of language and communication. Damage or dysfunction in any of these regions can result in language deficits or communication difficulties.

Vast body of research is available concerning the use of machine learning to diagnose Alzheimer's disease from voice recordings (see review papers [Pul20, Mar21, Vig22, Tha21, Yan22, Hec22]). There is no unified framework for testing and comparing those machine learning models, because research is done on vastly different datasets – both in terms of their statistical characteristics (the dataset size, the number of men and women, the number of healthy people and people in various stages of Alzheimer's disease), as well as their content and quality. There are several types of recordings used for diagnosing health disorders (including Alzheimer's disease):

1. longer recordings with spontaneous speech (typically pleasant stories, recollections of pleasant events, conversations) [Lop15a, Nas18],
2. longer recordings with recited parts of a chosen literary text [Mei14, Mar17],
3. longer recordings with answers to predefined inquiries (e.g. animal naming,, counting down, neuropsychological tests, sentence repeating, "Cookie theft" picture description task) [Ami22, Bal21, Kön18, Lag20, Per22, Xue21],
4. short recordings of a single vowel spoken by the subject (e.g. /a/, /e/, /u/).

Generating the first three types of recordings requires a higher level of comprehension from the subject, hence such recordings gather more information about a patient's intellectual state. It's no surprise then that the best results are obtained from models which use linguistic features (e.g. misspoken words, wrong words in the given context, semantic and phonetic errors) and speech features (e.g. articulation rate, duration of pauses, speech rate, degree of voice breaks). Acoustic features are also used in most models. These typically include a subset of features related to the fundamental frequency, harmonics, formants, spectrum and cepstrum. Some predefined sets of variables, like ComParE, are often used (see [Wen13]). The best models trained on longer recordings and using all types of features achieve accuracies even around 95% (see e.g. [Lop15a, Lop15b, Ami22, AlH16, Lag20, Nas18, Tho20]). Models based solely on acoustic features typically perform worse than those using also linguistic and speech features – typically with accuracies around 70-80% (see e.g. [Mei14, Per22, Xue21]). The best result on acoustic features was obtained in [Nas18] – accuracy of 97.7%, however it was achieved on multiple carefully chosen 60-second recordings from a small sample of 30 healthy subjects and 30 subjects with Alzheimer's disease.

Models based on vowel utterances are very rare in literature. A method of diagnosing MCI based on vowel utterances analysis from longer recordings of subjects speaking Swedish was described in [The18]. In [Nag20] one of the models was trained and tested solely on features created on vowel utterances, however details concerning recording length were not given. The ROC AUC obtained in this paper for distinguishing healthy subjects from those with MCI or Alzheimer was 0.63.

Our literature review showed that short recordings of single vowel utterances have been used in diagnosing voice pathologies (see [Abd22] and references therein), but not for diagnosing Alzheimer's disease (with the possible exception of the paper mentioned above). Almost all research was done on longer recordings of type (1), (2) and (3). Unfortunately, obtaining longer responses requires more time of the patient and is language-dependent. Our research focuses on very short recordings (6 seconds) containing lan-guage-independent utterances of the sustained vowel /a/. This approach renders our method universal.

To obtain good results with acoustic features based on very short recordings one has to put emphasis on feature selection based on physical understanding of the underlying processes. We achieved that with the use of sound objects – a technology of describing sound spectrum in a more accurate way than the typical spectrum (see Section 3). Based on sound objects we created a set of 14 features describing every recording (see Section 4).

On such a set of features we tested six models – XGBoost, LightGBM, SVM, SVR, Single Feature Linear Regression Ensemble (SFLRE), MultiLayer Perceptron (MLP) – against four scenarios:

1. healthy vs MCI,
2. healthy vs MCI and Alzheimer's,
3. healthy vs Alzheimer's,
4. MCI vs Alzheimer's.

We were able to obtain promising results: ROC AUC equal to 0.85 for the first scenario, 0.84 for the second, 0.87 for the third, 0.75 for the fourth (see section 5 for details). The accuracy was 0.76, 0.77, 0.76, 0.68, respectively.

## 2 | DATASET

The study included 300 individuals. Among them 90 were diagnosed with cognitive function disorders, comprising 64 females and 26 males. They were under the care of the Department of Psychiatry's at the Wroclaw Medical University. The control group consisted of 210 people – 176 females and 34 males.

All participants underwent a standard diagnostic procedure, which included a medical interview and laboratory tests to rule out other conditions that could cause cognitive functioVn disorders. Each patient also underwent neuroimaging, either through a computerized tomography (CT) scan or magnetic resonance imaging (MRI). In cases where there were contraindications for MRI, a CT scan was performed instead. Additionally, all participants underwent standard neuropsychological assessments. All patients were checked by MMSE and CDR tests and conducted interviews.

Following the diagnostic process based on ICD-10 criteria, the study confirmed the following diagnoses among the participants:
- Mild cognitive impairment (F06.7) was confirmed in 34 individuals.
- Alzheimer's disease (F00.2) was confirmed in 56 individuals.

The control group consisted of 210 individuals who were of an appropriate age and had no cognitive function disorders. The inclusion criteria for the study were as follows:

1. Diagnosis of mild cognitive impairment or dementia according to ICD-10 classification.
2. Exclusion of other psychiatric disorders.
3. A mental state that allowed for conscious consent to participate in the study.
4. No significant abnormalities in somatic health, as assessed through clinical examination and routine laboratory tests.

Exclusion criteria included:

1. Current severe somatic illnesses, especially circulatory system disorders, hypertension, liver or kidney dysfunction, malignancies, and metabolic diseases.
2. Past or present infectious diseases (such as brucellosis, Lyme disease, AIDS) and venereal diseases (syphilis, gonorrhea).
3. Neurological disorders, seizure disorders, or a history of severe head injuries.
4. Abuse or dependence on psychoactive substances or alcohol.
5. Visual or hearing impairments that hindered the execution of neuropsychological tests and recording procedures.

34 persons were excluded from the study due to the quality of the voice recordings. In the end the final dataset used for the Machine Learning models consisted of 266 observations, among which 186 (~70% of the dataset) were considered healthy, 46 were diagnosed with Alzheimer's and 34 with MCI. Among the 266 people, 215 were females and 51 males. While the share of MCI-diagnosed individuals was roughly the same in both groups (13.7% vs 12.6%), the share of people with Alzheimer's was twice as big among males than females (29.4% vs 14.4%).

**Table 1.** Counts of patients depending on their gender and diagnosis.

|  | Female | Male | All |
|---|---|---|---|
| **Alzheimer's** | 31 (14.4%) | 15 (29.4%) | 46 (17.3%) |
| **MCI** | 27 (12.6%) | 7 (13.7%) | 34 (12.8%) |
| **Healthy** | 157 (73.0%) | 29 (56.9%) | 186 (69.9%) |
| **Sum** | 215 (100%) | 51 (100%) | 266 (100%) |

The youngest person in the group was 26 years old while the oldest one was 88 years old. The mean age of the group was 60 years. For 38 people the age was unknown. The youngest person with a diagnosed MCI was 37 years old, while the oldest one was 78 years old. Similarly, among people with diagnosed Alzheimer's, the age started at 56 years and ended at 88 years. Looking at the histogram it can be seen that the majority of unhealthy individuals are in the older age groups.

**Fig 1.** Distribution of patients depending on their age and their diagnosis.

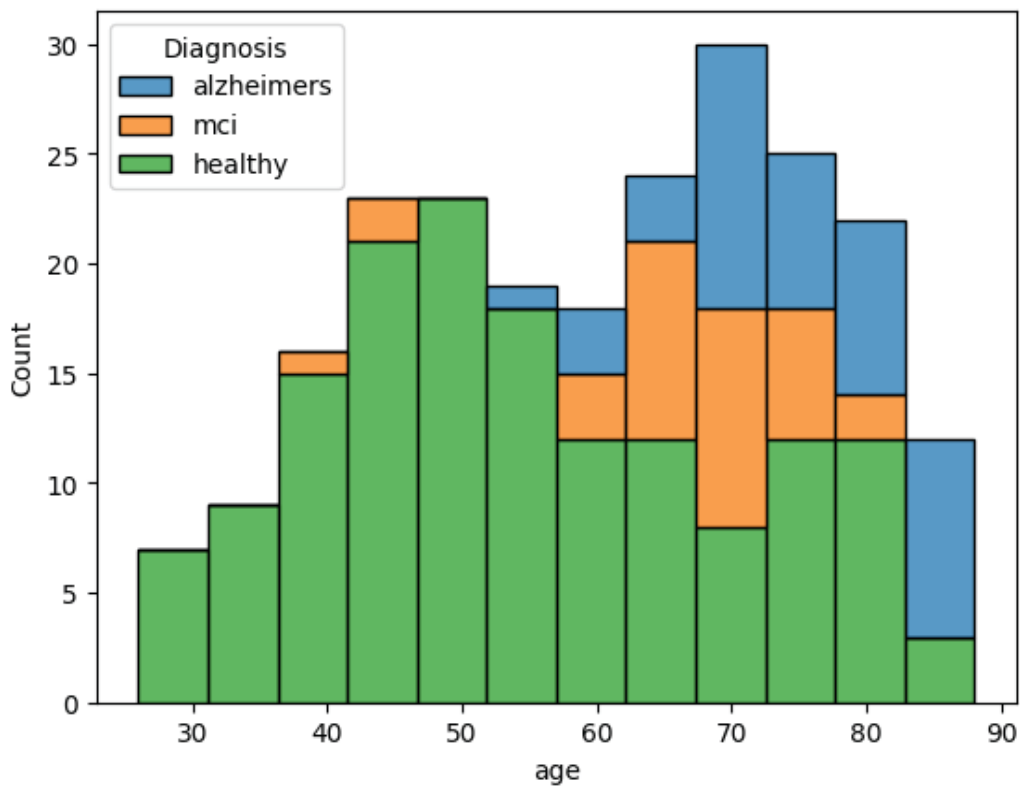

## 3 | SOUND OBJECTS TECHNOLOGY

Sound object technology introduces a new way of recording an acoustic signal.

Each sound object represents a sinusoidal acoustic signal with a slowly changing amplitude and a single slowly wavering frequency, characterized by phase continuity.

A sound object is recorded as a sequence of points represented by three parameters: position (time), amplitude, and frequency. Additionally, each object has a calculated initial phase, thanks to which its amplitude, frequency, and phase can be calculated for each time moment and component of the acoustic signal.

The information stored in sinusoidal components depends on their frequency, therefore the distance of the points depends on the frequency of the object.

Measurements show that with a point distance of approximately 2 periods, the reproduction of the acoustic signal by objects reaches 99.5% (measured in terms of pointwise energy difference between the original and the reproduced signal), and at the same time reduces the number of parameters needed to record all signal components.

The recorded acoustic signal is saved as a sequence of subsequent measurements made using microphones at intervals of 22 050 or 44 100 samples per second.

The basic task of the technology is to isolate all components from the recorded signal without distorting their parameters. The designated objects must reproduce the change in parameters of real signals by following them.

To determine sound objects, a Bank of zero-phase filters (not distorting the phase) with a logarithmic frequency distribution of 48 filters per octave was used.

This means that in each interval in which the frequency doubles there are 12 semitones, and in each semitone there are 4 filters.

To process the human voice in the band from 64 Hz to 10,000 Hz, 350 filters are needed. The filter bank creates a composite spectrum of the signal, producing 62 MB of data per second.

Measuring the value of the spectrum determines the places where the energy is maximal. The phase difference in successive samples of the spectrum determines the frequency of the component signals. The places where the measured frequency corresponds to the rated frequency of the filter are indicated by subsequent points of sound objects. If there is no match between the phase difference and the frequency, it means that the signal component has been disturbed or has ended. At this point, the sound object will end and a new object may be created with a new starting phase.

The presented acoustic signal vectorization technology has been patented (see [Plu18]).

## 4 | SOUND OBJECT FEATURES AS VOICE BIOMARKERS

Sound object technology is particularly well suited to measuring the characteristics of a single person's voice. When recorded by one person, sound objects create parallel structures and are not disturbed by harmonic structures from other sources.

Figure 2 shows three examples of the analyzed voice from a healthy person (left column), a person diagnosed with mild cognitive impairment – MCI (middle column), and a person diagnosed with Alzheimer's disease (right column).

Fig. 2.a shows an acoustic signal – a fragment of the recording of the sound aaaa.... In a healthy person, the signal periods are very even and repeatable. In a person with MCI disorder, subsequent cycles of the acoustic signal are distorted. Harmonic components are shifted. In a sick person, the distortions are more severe. The fundamental component is fragile. The signal is twice as rare. It is difficult to visually capture the differences from the audio signal alone, so we use sound object technology.

Fig. 2.b shows the complex spectrum of the above signal. In a healthy person, the spectrum is very sharp. Energy is concentrated in harmonic structures. The dorsal edge is smooth and straight. In a person with MCI disorders, one can notice areas where the sound vibrates in amplitude – the wrinkled dorsal edge (the depth dimension in the chart), as well as in frequency (the vertical dimension in the chart).

It is difficult to distinguish an even edge of the spectrum in a sick person. The back seems swollen from shaking.

Fig. 2.c shows sound objects that are the result of the analysis of the spectrum ridge while maintaining phase continuity. When the phase of the signal in the next sample does not correspond to the frequency of the spectrum, the object is interrupted and a new object is created with a new initial phase, thanks to which there remains a trace that there was a disturbance or a temporary loss of sound emission control in the analyzed signal.

Fig. 2.d shows the remaining non-harmonic objects (the energy of each of which is > 1% of the energy of the entire signal) in purple, and noise signals: low noise up to 200 Hz in gray, medium noise up to 2000 Hz in green and high noise in blue.

Fig. 2.e contains all objects creating harmonic structures, selected based on frequency analysis. At the bottom, the fundamental component corresponding to the glottal frequency is marked in red. Strong harmonic components are marked in brown, each of which has an energy exceeding 5% of the energy of the entire signal. Other objects belonging to harmonics are marked in yellow.

Fig. 2.f shows an acoustic signal composed only of harmonic components, showing how accurately the recorded signal can be reproduced using only harmonic components.

**Fig 2**  Stages of acoustic signal analysis

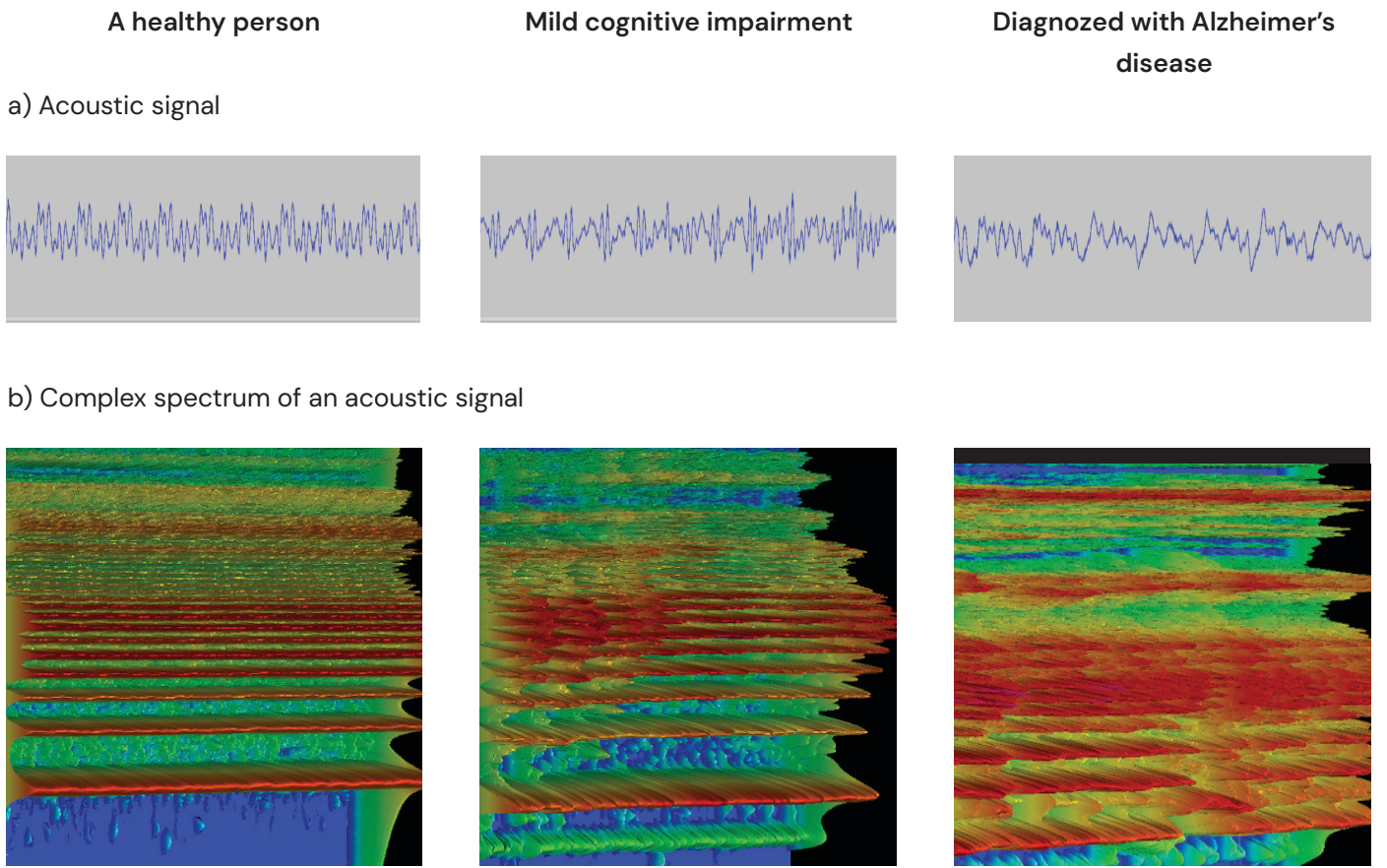

a) Acoustic signal

b) Complex spectrum of an acoustic signal

Columns: A healthy person | Mild cognitive impairment | Diagnozed with Alzheimer's disease

c) Sound objects

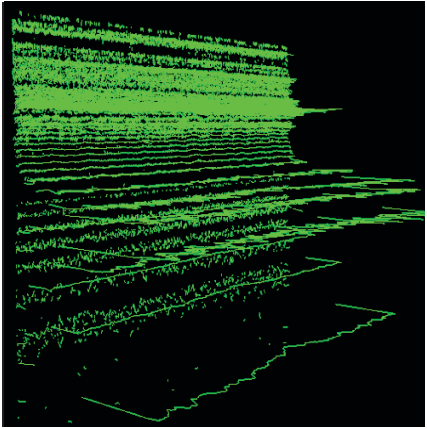
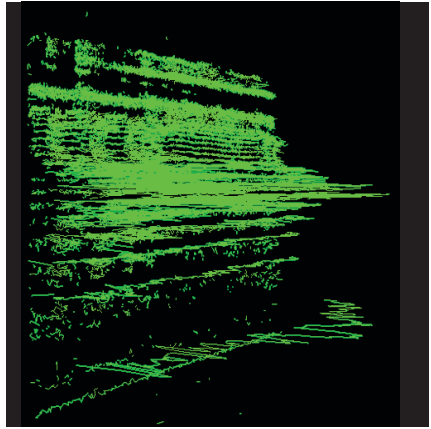
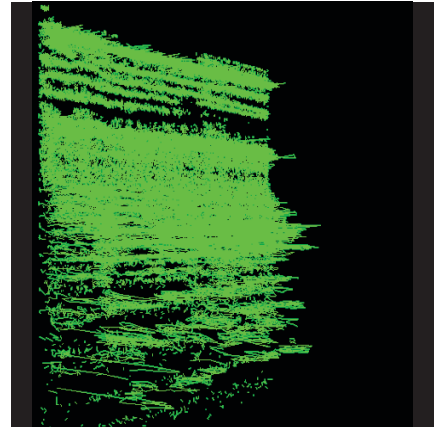

d) Noisy and inharmonic objects

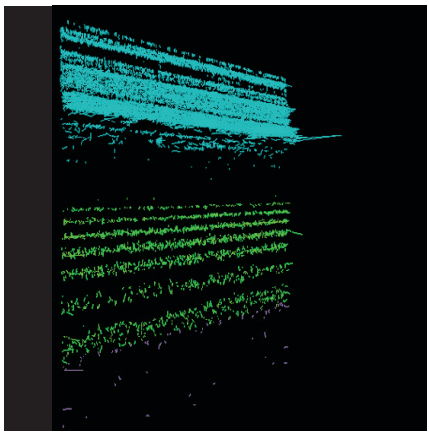
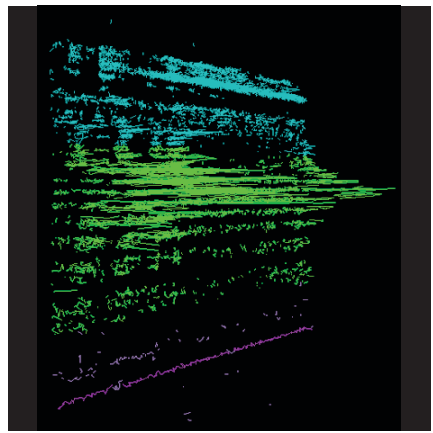
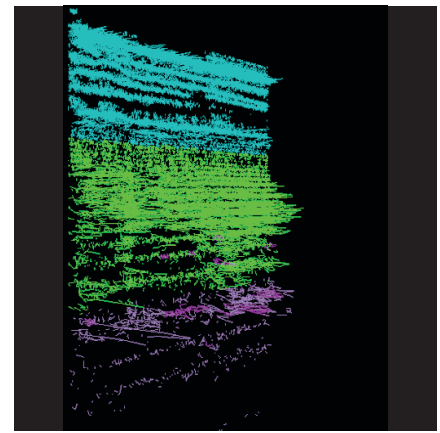

e) Harmonic components

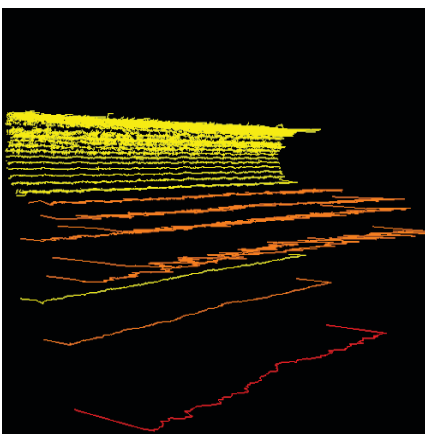
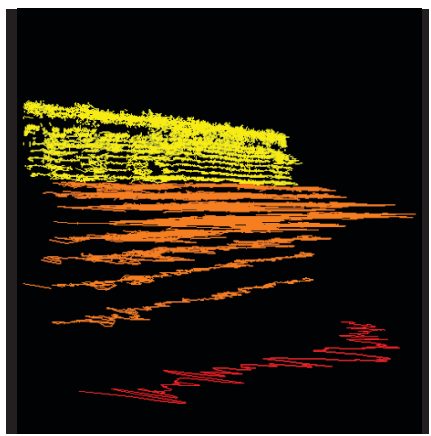
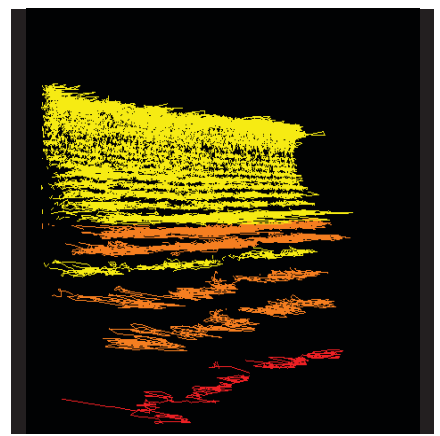

f) An acoustic signal composed of harmonic components

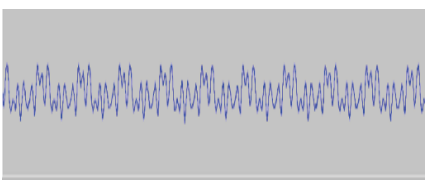
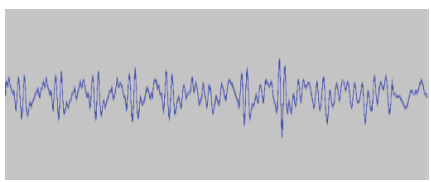
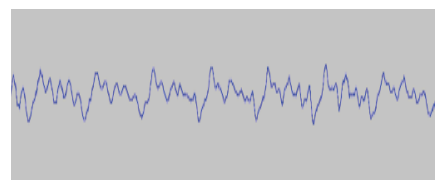

Having precisely measured parameters of sound objects that change over time, it is possible to assess the quality of the voice emitted by the examined person much more precisely than visually.

## The first stage of the analysis – measurement of objects

In the first stage of the analysis, changes in amplitude and frequency parameters over time are counted for all sound objects. The parameters are measured in local intervals (e.g. 5 times per second) and global intervals (for the entire recording).

For the amplitude and frequency of each object the following are calculated:

- **Average value**
  
  **The average** value of the object's amplitude represents the strength of the sound component. The share of the object's energy (proportional to the square of the average amplitude) in the energy of the entire signal determines the role of this object in the entire recording.
  
  **The average frequency value** allows, in the next stage, to classify the object into a group of signals such as:
  
  **harmonic signals** – whose frequencies are multiples of the fundamental frequency;
  
  **inharmonic signals** - with a frequency not corresponding to a multiple of the fundamental frequency, but strong enough to play an important role in the recording (defined as having at least 1% of the energy of the entire recording);
  
  **noise signals** – other objects.

- **Standard deviation**
  
  **The standard deiation of the amplitude** describes how much the amplitude varies over time. How the volume of the voice rises and falls. Local measurements allow us to distinguish whether changes in voice intensity result from the interpretation and expression of emotions or are caused by a lack of control over the voice
  
  **The standard deviation of frequency** describes how strongly the pitch changes. And here, local measurements allow us to determine whether the change in voice pitch is due to intonation or indicates a disease state.

- **Vibrations (total variance)** in which the sums of the absolute values of differences of adjacent points of a sound object are measured.
  
  **Amplitude vibrations – Shimmer** describes how strong short changes are. How the signal amplitude flickers. During speech, the sound should not vibrate quickly, and if it does, it usually indicates problems with the control of the speech system.
  
  **Frequency vibration – Jitter** describes how strong the frequency vibrations are. Strong, fast frequency oscillations often signal abnormalities in the control of the speech system.

- **The slope coefficient** is the slope of a linear regression model fit to a given parameter.
  
  **Amplitude Slope – AmpLean** describes a trend or slow change in amplitude. A parameter value close to zero indicates good control over the voice. Measuring the parameter in subsequent time intervals may help in describing emotions.
  
  **Frequency Slope – FreqLean** describes the tendency or slow change in the frequency pitch of a measured signal. When examining a patient's health, the value of the parameter should be low. The parameter measurement in the following sections can monitor intonation.

- **Quantitative parameters of objects**
  
  When determining parameters, the number of points in the object, the length of the object and its energy are stored. These parameters are summed locally and globally to determine their total value in the analyzed intervals.

## The second stage of analysis – grouping objects

In the next stage, sound objects are grouped for further analysis.

All objects with an energy exceeding 1% of the energy of the entire recording or with a long duration exceeding 1 second are selected. This small (about several dozen) group of objects plays a decisive role in the recording.

By comparing the average frequencies of all selected objects, groups of objects are created with **a**

**common frequency divisor.** The group in which the sum of energy is largest will form the **harmonic group.** The lowest common divisor creates the **fundamental frequency F1.** Successive integer multiples of the fundamental frequency create subsequent harmonics from **the second harmonic F2** to the **twenty-third harmonic F23.** The higher harmonics are usually very weak and fragmented, and their frequencies are so close to each other that it is difficult to classify them.

The remaining objects selected and not separated into harmonics are classified as **non-harmonic (Sub Harmonics).**

All objects left so far, whose average frequency corresponds to the frequency of the harmonic group, will be assigned to the **fundamental group** and **subsequent harmonics.** When assigning objects to harmonics, the statistical parameters of the group are calculated based on the objects' parameters.

The group of **Strong Harmonics (Low Harm)** will include F1 and all harmonics whose total energy of the objects belonging to it is greater than or equal to 5% of the energy of the entire signal.

The remaining harmonics will form the group of **Weak Harmonics (High Harm).**

Objects not assigned to harmonics create **Noise**, in which objects are grouped into **Low Noise** – with a frequency **< 200 Hz, Medium Noise** with a frequency **from 200 Hz to 2000 Hz** and **High Noise** – with a frequency **above 2000 Hz.**

**The third stage of the analysis – determining phase relations for the Strong Harmonics**

- **Average value of the Harmonic Group Phase** – based on the amplitude and phase of sound objects included in the harmonic, the resultant harmonic phase is calculated at the time when the fundamental phase F1 has the value 0 – it is called harmonic shift. Several measurements are made within each time section and an average value is calculated from the measurements. The speech signal, and in particular the voiced sounds, consists of the sum of harmonic signals shifted among themselves in a characteristic way. In a healthy person, this shift is constant, thanks to which the graphical graph of the acoustic signal is repeatable with a period corresponding to the fundamental frequency. The repetitive phase shift is a characteristic of the speaker. If more research confirms that this feature is long-lasting, the average phase value could be used to identify people.
- **Phase standard deviation** – from the result of measuring the harmonic phase and calculating the average phase value, the dispersion of the measured phase parameters can be determined in a similar way as for amplitude and frequency. The phase standard deviation is more sensitive to ailments and problems with the brain's control of the speech system than frequency oscillations. Differences in the harmonic shift, and therefore in the appearance of the acoustic signal, can be easily observed by visually examining subsequent periods of the audio signal.
- **Phase Drift** – short-term, fast phase changes, calculated as the total variance of harmonic shifts. Notice that if all harmonic shifts are identical (signifying good voice control), then phase drift is zero. Drift can be observed after just a few measurements and is a clear sign of problems with speech control.

**The fourth stage of the analysis – measurement of quantitative parameters and energy distribution**

- **Sound Objects/Harm.** A characteristic feature of sound objects is phase continuity tracking, so when for some reason (e.g. due to disruption or problems with the control of the speech system) there is a quick change in the phase of the object, the object is closed and after a while a new object is created with a new phase. This means that an increased number of objects within a harmonic group is evidence of a disorder or disease state. In an ideal situation, each strong harmonic should be composed of one object.
- **Low Harmonics Energy.** The basic information contained in the voice of a healthy person is lo-

cated in the Low Harmonics (between 65% and 85% of the energy should be contained in these objects). In the voice of a healthy person, some of the energy should also be found in high harmonics and medium and high noise (the remaining 35 to 15%).

- **Sub Harmonics Energy.** Non-harmonic objects are not desirable. It is a wheezing, gurgling, sharp vibration that does not indicate disease, but usually occurs in large amounts in a person with dementia.
- **Harmonic to Noise Ratio.** Energy ratio of all harmonics to noise and non-harmonics. Optimally, for a healthy person, harmonics should contain much more energy, but a complete lack of noise is not good either, as it may indicate poor quality of the recording or the poor health of the subject. In a sick person, higher harmonics often disappear, the noise is weak and only a small number of harmonics and non-harmonics remain.
- **Harmonics Tilt** is the slope coefficient of the Low Harmonics energy distribution graph, calculated in a similar way to the Lean slope of amplitude and frequency, except that the parameters are the energy of the fundamental component F1 and the energy of the Strong Harmonics components. In a healthy person, the energy distribution in the sound aaa... should be relatively flat. In a person with speech difficulties, most of the energy is in the fundamental F1, which causes the slope to be negative, or vice versa, a lot of energy is concentrated in one harmonic, e.g. the 5th harmonic, and there is little in the others, which causes the line to tilt high.

**Table 2.** Comparison of calculated sound timbre parameters.

| Parameters | A healthy person | | | Mild cognitive impairmen | | | Diagnosed with Alzheimer's disease | | |
|---|---|---|---|---|---|---|---|---|---|
| | I k | M | III k | I k | M | III k | I k | M | III k |
| **Amplitude – percentage change in the average value** | | | | | | | | | |
| Standard deviation | 11.6 | **16.7** | 26.1 | 27.8 | **33.9** | 39.1 | 31.2 | **35.7** | 38.6 |
| Shimmer | 0.6 | **1.0** | 1.4 | 2.0 | **2.5** | 3.4 | 2.8 | **3.4** | 4.0 |
| Slope coefficient | 1.5 | **4.2** | 7.1 | 8.1 | **10.7** | 19.5 | 17.5 | **22.0** | 28.1 |
| **Frequency – percentage change in the average value** | | | | | | | | | |
| Standard deviation | 0.8 | **1.0** | 1.5 | 1.3 | **1.6** | 2.5 | 1.6 | **2.4** | 3.1 |
| Jitter | 0.03 | **0.04** | 0.07 | 0.09 | **0.10** | 0.14 | 0.13 | **0.16** | 0.19 |
| Slope coefficient | 0.01 | **0.08** | 0.2 | 0.2 | **0.6** | 0.8 | 0.75 | **1.9** | 2.5 |
| **Phase – shift of the phase of the harmonic components relative to the zero phase of the fundamental component – in radians** | | | | | | | | | |
| Standard deviation | 0.23 | **0.63** | 1.14 | 1.05 | **1.67** | 2.01 | 1.63 | **1.95** | 2.03 |
| Vibration – Drift | 0.15 | **0.37** | 0.86 | 1.05 | **1.61** | 2.35 | 1.41 | **1.99** | 2.40 |
| **Average number of objects per strong harmonic component** | | | | | | | | | |
| OBD/Harm | 1 | **3** | 7 | 9 | **12** | 16 | 11 | **18** | 25 |
| **Number of non-harmonic objects with energy > 1% of the total signal energy** | | | | | | | | | |
| SubHarm | 0 | **2** | 6 | 2 | **5** | 11 | 5 | **10** | 12 |
| **Percentage distribution of energy in components** | | | | | | | | | |
| ELowHarm | 87.2 | **83.5** | 78.9 | 78.5 | **74.2** | 68.3 | 71.4 | **57.8** | 49.6 |
| ESubHarm | 0.0 | **0.1** | 0.5 | 0.5 | **1.6** | 3.7 | 2.1 | **3.9** | 7.6 |
| **Ratio of harmonic energy to noise and non harmonic** | | | | | | | | | |
| HNR | 14.7 | **23.3** | 33.8 | 5.4 | **7.1** | 11.5 | 2.1 | **3.5** | 5.5 |
| **The slope of the energy distribution line between strong harmonics** | | | | | | | | | |
| FqTilt | 0.56 | **1.46** | 3.37 | 0.64 | **1.41** | 2.54 | 0.57 | **2.01** | 6.13 |
| **Number of respondents** | | 186 | | | 46 | | | 34 | |

Continuing the development of the application towards tracking changes in the above-mentioned parameters in time sections, it seems possible to build and track other, e.g. prosodic parameters of sound timbre, such as melody, intonation, melodiousness, accent, dynamic force, rhythmicity, etc.

The presented method of determining sound tone parameters was described in the patent application [Plu23].

Table 1 shows examples of measuring the above-described sound timbre parameters based on the recording of the sound aaa... for 310 people.

The people were classified by doctors into three categories – a healthy person, a person with mild cognitive impairment and a person diagnosed with Alzheimer's disease.

All recordings were vectorized using Sound Object Technology. Then, the calculated sound objects underwent the described four stages of analysis.

For the global parameters of each group of patients, 14 measurement results were ranked and Ik – First quartile, M – Median and IIIk – Third quartile were determined. The results are presented in the table 2.

## 5 | MODELS

The following models has been tested in this research:
- Boosting:
  ◇ eXtreme Gradient Boosting (XGB),
  ◇ Light Gradient Boosting Machine (LGBM),
- Support Vector Machines (SVM):
  ◇ Epsilon-Support Vector Regression (SVR),
  ◇ C-Support Vector Classification (SVC),
- Single Feature Linear Regression Ensemble (SFLRE),

- MultiLayer Perceptron (MLP).

For the Boosting algorithms the XGBoost and LightGBM libraries were used respectively, specifically the xgboost.XGBClassifier and lightbgm.LBGMClassifier classes.

For the Support Vector Machines the scikit-learn library was used, specifically the sklearn.svm.SVR and sklearn.svm.SVC classes. Also the Single Feature Linear Regression Ensemble used the scikit-learn library, specifically the sklearn.linear_model.LinearRegression class.

For the Neural Networks the PyTorch library was used and a tailor-made class was used.

In the MultiLayer Perceptron the output layer used a logistic sigmoid function. For the input and hidden layers the ReLU activation function was utilized. Each architecture also used a dropout for regularization. The number of hidden layers varied from 0 to 4 and was dependent on the target variable and the Optuna objective which was being optimized.

The concept behind the Single Feature Linear Regression Ensemble was to use linear regression to forecast the target variable using one feature at a time. Since there were 16 features, this resulted in 16 models. The forecasts from these models were averaged and the final result was treated as probability.

## 6 | EXPERIMENTS

In total a few hundred different experiments were conducted. Each experiment was verified by training 10 models by using a different random seed (if a model accepted a random seed as a parameter) and, using the same random seed, each model was trained and tested using 5-fold stratified cross-validation. In total 50 different models were trained and their resulting metrics were averaged to achieve the final results.

There were 16 distinct features which were used in the models:
- 14 variables described in Section 4,
- gender [bool] – female or male,
- age [int] – the age of the person when the recording took place.

In boosting, SVMs and NNs all of the above features were used and in MOF-LR all features bar gender and age. No additional feature engineering or feature selection was conducted (experiments with best k features did not show any improvement in results).

For the data points where age was missing, the average age of all observations was imputed.

There were 4 binary target variables verified and depending on the target variable the dataset had a different number of observations:
- **healthy vs MCI** (0 – healthy; 1 – MCI), total: 220 observations,
- **healthy vs MCI /Alzheimer's** (0 – healthy; 1 – MCI/Alzheimer's), total: 266 observations,
- **healthy vs Alzheimer's** (0 – healthy; 1 – Alzheimer's), total: 232 observations,
- **MCI vs Alzheimer's** (0 – MCI; 1 – Alzheimer's), total: 80 observations.

For all target variables the probability threshold was set to the share of 1s in the dataset. Therefore the threshold was different depending on the dataset and CV set.

Depending on the model the feature scaling technique which provided the best results was used:
- for Boosting and SFLRE – no scaler,
- for SVM – Standard Scaler,
- for ANN – Quantile Transformer.

While Boosting and Linear Regression were algorithms for which feature scaling didn't bring noticeable improvements in results, SVMs gave better results when the features were standardized. This can be attributed to the fact that in SVMs the distances between points matter, since we look for a decision boundary which maximizes the distances between the nearest point of each class. For Neural Networks scaling is highly recommended since it helps with the gradient descent optimization. The Quantile Transformer transforms the data in such a way that values of every feature are uniformly distributed. This makes features with different value ranges comparable with each other and allows for picking subtleties in data concentrated around a single point even in the case of very far outliers appearing in the data.

Every model except for SFLRE went through a hyperparameter optimization process using the Optuna library. There were 600 trials conducted for each model. The CV groups used in the Optuna trials were different (inner CV) than the groups which were used to calculate the final results (outer CV). The trials were repeated for 2 different objectives:
- maximizing the ROC AUC (ROC),
- maximizing the sensitivity and precision sum (SPS).

In half of the cases one objective gave better results and in the other half the other objective. Specifically when the target variable was "healthy vs MCI" maximizing the ROC AUC gave higher results for all models while for other target variables usually maximizing the sum of sensitivity and precision was more optimal. In terms of models the MLP gave better results in 3 out of 4 cases when using the ROC AUC as the objective. Similarly, when using SVR, in 3 out of 4 cases better results were achieved while using the sensitivity and precision sum. In other models there was no clear winner when it came to the objective.

The results show that no single model achieved the best results. There were however models which achieved good results more often. Specifically LGBM and XGB had the 15 best results across all target variables and metrics, while SFLRE had 0. Also the MLP fared well (5 best results) specifically when Alzheimer's was one of the possible states. The SVR had 3 out of 4 best results in terms of sensitivity (close to 1) however that also resulted in a specificity result lower than the other models. This suggests that the SVR most likely didn't learn about the underlying dependencies and in most cases simply forecasted the value 1.

**Table 3.** Average metrics for all models and target variables. Results are denoted as inner / outer CV. Bold results are the highest for a given metric and target variable.

| Target variable | Model | Optuna target | ROC AUC | SENSITIVITY | SPECIFICITY | ACCURACY | F1 |
|---|---|---|---|---|---|---|---|
| healthy vs MCI | SFLRE | N/A | 0.74 / 0.74 | 0.72 / 0.73 | 0.65 / 0.65 | 0.66 / 0.66 | 0.40 / 0.40 |
|  | LGBM | ROC | 0.86 / **0.85** | 0.83 / 0.83 | 0.74 / 0.74 | 0.76 / **0.76** | 0.52 / **0.52** |
|  | XGB | ROC | 0.85 / 0.84 | 0.83 / 0.82 | 0.75 / **0.75** | 0.76 / **0.76** | 0.52 / **0.52** |
|  | SVR | ROC | 0.75 / 0.74 | 1.00 / **1.00** | 0.00 / 0.00 | 0.15 / 0.15 | 0.27 / 0.27 |
|  | SVC | ROC | 0.74 / 0.73 | 0.63 / 0.65 | 0.72 / 0.71 | 0.70 / 0.70 | 0.40 / 0.40 |
|  | MLP | ROC | 0.74 / 0.72 | 0.80 / 0.79 | 0.49 / 0.44 | 0.53 / 0.49 | 0.36 / 0.33 |
| healthy vs MCI & Alzheimer's | SFLRE | N/A | 0.80 / 0.79 | 0.75 / 0.75 | 0.71 / 0.72 | 0.72 / 0.73 | 0.62 / 0.62 |
|  | LGBM | SPS | 0.84 / 0.80 | 0.85 / 0.86 | 0.74 / 0.74 | 0.77 / **0.77** | 0.69 / **0.69** |
|  | XGB | SPS | 0.86 / **0.84** | 0.96 / **0.95** | 0.63 / 0.63 | 0.73 / 0.72 | 0.68 / 0.67 |
|  | SVR | SPS | 0.80 / 0.81 | 0.95 / 0.94 | 0.52 / 0.52 | 0.65 / 0.65 | 0.62 / 0.62 |
|  | SVC | ROC | 0.81 / 0.81 | 0.74 / 0.73 | 0.74 / **0.75** | 0.74 / 0.74 | 0.63 / 0.62 |
|  | MLP | SPS | 0.81 / 0.81 | 0.90 / 0.86 | 0.63 / 0.66 | 0.71 / 0.72 | 0.66 / 0.65 |
| healthy vs Alzheimer's | SFLRE | N/A | 0.85 / 0.85 | 0.81 / 0.81 | 0.70 / 0.70 | 0.72 / 0.72 | 0.54 / 0.53 |
|  | LGBM | ROC | 0.88 / **0.87** | 0.83 / 0.83 | 0.75 / **0.74** | 0.76 / **0.76** | 0.58 / **0.58** |
|  | XGB | SPS | 0.85 / 0.82 | 0.86 / 0.79 | 0.75 / 0.73 | 0.77 / 0.74 | 0.60 / 0.55 |
|  | SVR | SPS | 0.86 / 0.86 | 0.97 / **0.97** | 0.56 / 0.57 | 0.64 / 0.65 | 0.52 / 0.53 |
|  | SVC | SPS | 0.86 / 0.86 | 0.90 / 0.91 | 0.65 / 0.61 | 0.70 / 0.67 | 0.55 / 0.54 |
|  | MLP | ROC | 0.88 / **0.87** | 0.89 / 0.86 | 0.71 / 0.72 | 0.75 / 0.75 | 0.59 / **0.58** |
| MCI vs Alzheimer's | SFLRE | N/A | 0.62 / 0.62 | 0.51 / 0.51 | 0.63 / 0.65 | 0.56 / 0.57 | 0.55 / 0.57 |
| | LGBM | SPS | 0.70 / 0.73 | 0.69 / 0.68 | 0.63 / 0.65 | 0.66 / 0.67 | 0.70 / 0.69 |
| | XGB | ROC | 0.75 / **0.75** | 0.64 / 0.67 | 0.62 / 0.64 | 0.64 / 0.66 | 0.66 / 0.68 |
| | SVR | SPS | 0.68 / 0.68 | 0.76 / **0.76** | 0.44 / 0.41 | 0.63 / 0.61 | 0.70 / 0.69 |
| | SVC | SPS | 0.70 / 0.70 | 0.68 / 0.63 | 0.59 / 0.63 | 0.64 / 0.63 | 0.67 / 0.65 |
| | MLP | ROC | 0.75 / 0.74 | 0.68 / 0.69 | 0.66 / **0.66** | 0.67 / **0.68** | 0.70 / **0.71** |

The average ROC AUC varied from 0.62 to 0.87 depending on the target variable and model. The results of specific CV tests varied from 0.40 to 0.98 which shows that the specific observations which were used for the training process and then the testing had a significant impact on the final result. Specifically when the target variable was "MCI vs Alzheimer's" all models apart from the MLP had multiple cases of ROC AUC below 0.5, meaning that in those cases the models forecasted worse than a random draw.

**Plot 1.**　　Whisker plots which show the ROC AUC value distribution for all 50 results for each model and target variable

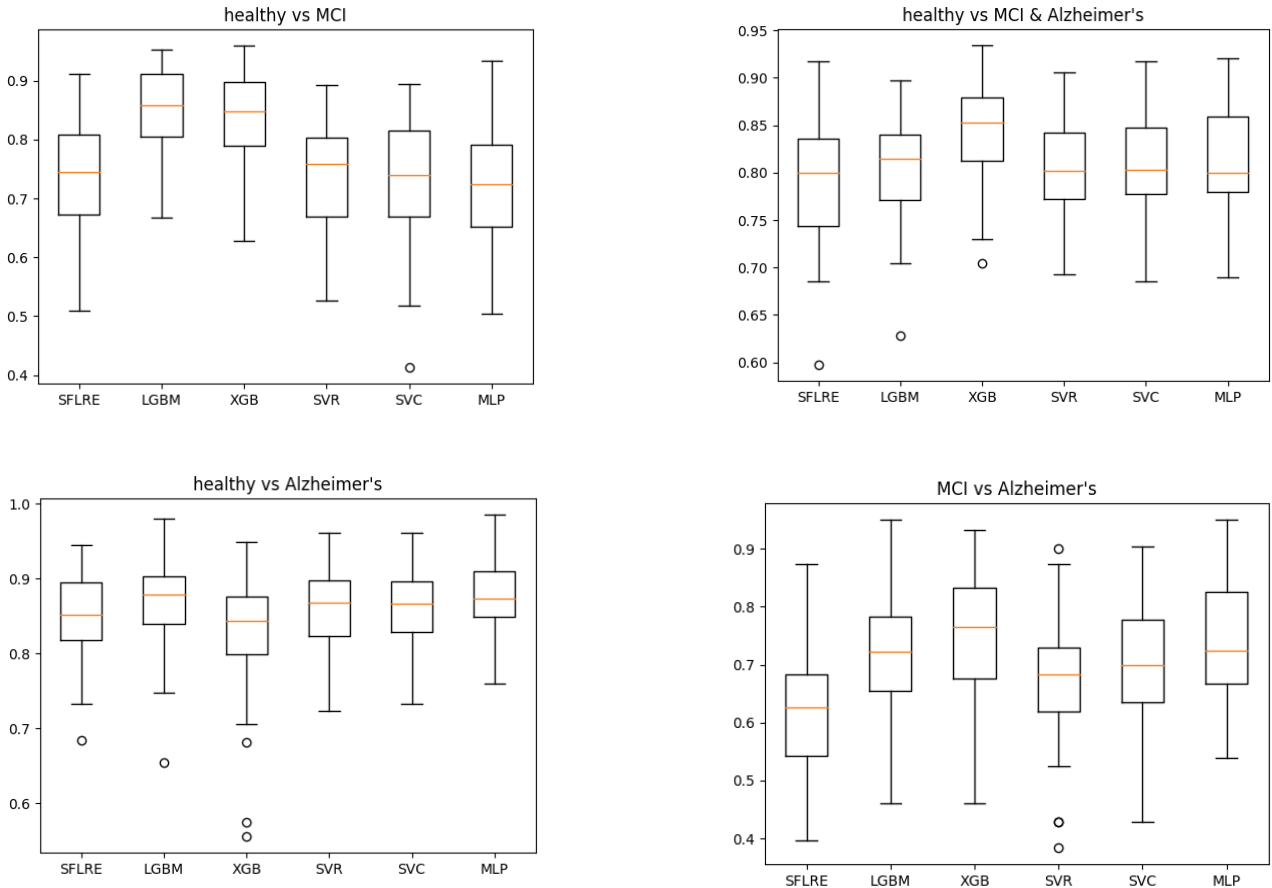

# 7 | CONCLUSION

The article presents a study on the use of machine learning models to differentiate between healthy individuals, those with mild cognitive impairment (MCI), and diagnosed with Alzheimer's disease based on voice biomarkers. The final dataset used for machine learning models consisted of 266 observations, with a distribution of 186 healthy individuals, 46 diagnosed with Alzheimer's, and 34 with MCI. This sizable dataset provided a solid foundation for training and testing the models. The age of participants ranged widely, from 26 to 88 years old, with the mean age being 60 years. This wide age range demonstrates the applicability of voice biomarker analysis across a broad spectrum of the adult population. Among the 266 individuals, 215 were females and 51 males. The share of Alzheimer's diagnosis was twice as high among males (29.4%) compared to females (14.4%), indicating potential gender-related differences in the prevalence or detection of the disease.

The study demonstrates that sound object technology, which analyzes the characteristics of a person's voice, is an effective tool in distinguishing between healthy individuals, those with MCI, and Alzheimer's patients. The diagnostic accuracy was high. The machine learning models used for voice analysis demonstrated statistically significant diagnostic accuracy. Models XGB and LGBM achieved ROC AUC (Receiver Operating Characteristic Area Under the Curve) scores as high as 0.86 and 0.84, respectively, in distinguishing between healthy individuals and those with cognitive impairments. Models also showed high sensitivity (up to 0.96 for XGB in detecting MCI and Alzheimer's) and specificity (up to 0.75 for LGBM in the same context), indicating their effectiveness in correctly identifying patients with cognitive disorders and those without.

We also found significant differences in key voice parameters among healthy individuals, those with MCI, and Alzheimer's patients. Parameters like amplitude (shimmer), frequency (jitter), and along with the decreasing harmonic quality of the voice (HNR), correlate with the severity of cognitive impairment. Slope coefficients showed marked variation across these groups. Among others the following patterns have been observed:

- Healthy individuals: Shimmer values ranged around 0.6 to 1.4, indicating relatively stable amplitude. Jitter values were lower, typically around 0.03 to 0.07, indicating stable frequency. Higher HNR values from 14.7 to 33.8, indicating a clearer and more harmonic voice quality.
- MCI patients: Shimmer values were higher, ranging from 2.0 to 3.4, showing increased amplitude variability. Jitter values increased to 0.09 to 0.14, showing more frequency variability. Reduced HNR values from 5.5 to 11.5, suggesting a decrease in voice clarity and harmonic quality.
- Alzheimer's patients: Shimmer values further increased to 2.8 to 4.0, suggesting even greater amplitude fluctuations. Jitter values were even higher, ranging from 0.13 to 0.19, indicating significant frequency instability. Further reduced HNR values from 2.1 to 5.5, indicating more pronounced deterioration in voice quality.

These numbers underscore the potential of voice biomarkers as a reliable, non-invasive, and accessible tool for early detection cognitive changes. But to validate the effectiveness of voice biomarkers in monitoring disease progression, long-term studies are required. The next step in our research is to use features based on sound objects calculated for fragments of the recording and create a machine learning model (potentially deep learning) which would spot deterioration in a patient's speech control based on the analysis of changes in speech characteristics within the recording. The automated nature of machine learning analysis allows for the processing of large volumes of data quickly. This makes it suitable for mass screening programs, which could be particularly useful in identifying at-risk populations. Widespread use of voice analysis could contribute to large-scale data collection, and can inform public health strategies and research.

This study will have important implications for the research on inexpensive and efficient screening methods for dementia. The ability to differentiate between healthy, MCI and Alzheimer's through voice analysis offers significant potential for early diagnosis. Early detection is crucial for the effective management and treatment of these conditions. The use of voice biomarkers for diagnosing cognitive disorders represents a non-invasive and easily accessible method as in our study we can use smartphones to record voice samples. This is especially important for early screening, as traditional diagnostic methods like laboratory tests, neuroimaging or extensive neuropsychological testing can be costly, time-consuming, and sometimes invasive. Voice analysis can be conducted remotely and with minimal equipment, making it accessible to a wider population, including those in rural areas or low-reource settings. This increases the potential for early detection among a broader demographic. Currently, efforts are being made to personalize treatment. As machine learning models identify specific voice patterns associated with different stages of cognitive impairment, treatment approaches can be more personalized. Healthcare providers can tailor interventions based on the individual's specific condition and progression rate. With earlier diagnosis, interventions, whether pharmacological or lifestyle-based, can be initiated sooner, potentially slowing the progression of the disease, individuals can maintain independence and cognitive function for longer. Changes in voice biomarkers can provide insights into the effectiveness of treatments and the progression rate of the disease, allowing for timely adjustments in treatment plans.

Acceptance of psychiatric treatment varies across cultures. This method could help in reducing the stigma associated with psychiatric compliance, cognitive testing and encourage more people to seek early diagnosis. Due to the non-invasive and simple nature of voice recording, individuals might be more inclined to participate in early screening. This can lead to greater engagement in proactive health management.

Unlike most methods developed in this research area, the proposed method requires only short recordings of language-independent speech – sustained vowel /a/ which is similar in different languages. Hence it should improve global accessibility of this method as this technology can be made applicable in diverse linguistic and cultural settings. Recording voice samples is relatively simple and does not require extensive training. This means that a wide range of healthcare providers, from specialists to general practitioners, can use the technology.

Voice analysis from our study can be easily integrated into telemedicine platforms, allowing for remote monitoring and assessment. This is particularly beneficial for patients who are unable to travel to healthcare facilities regularly, such as the elderly or those living in rural areas. Compared to traditional diagnostic methods like MRI or CT scans, voice analysis is much more cost-effective. This could lead to its adoption as a first-line screening tool, reducing the need for more expensive and invasive tests.

The study points to several future directions that can enhance the use of voice biomarkers and machine learning in the diagnosis and monitoring of cognitive disorders. These future directions aim to address current limitations and expand the utility of this technology. The accuracy of machine learning models is highly dependent on the quality and quantity of the data they are trained on. Inconsistent or poor-quality voice recordings can lead to inaccurate analyses. The technology's efficacy is subject to the limitations of the recording devices used and the acoustic environments in which recordings are made. Background noise, microphone quality, and recording settings can all impact the quality of the data collected. Additionally, a large and diverse dataset is essential to train robust models that can generalize well across different populations. The study showed that different machine learning models yielded varying results. This inconsistency can be a challenge in standardizing the approach for clinical use. Crucial for the technology's effectiveness should be finding the most reliable and accurate model or combination of models. At the same time, addressing ethical concerns related to data privacy is very important. Developing protocols for secure data handling and ensuring transparency in how voice data is used will be important for public trust and ethical compliance. Conducting rigorous clinical trials and validation studies is necessary to establish the clinical efficacy of voice biomarker analysis. These studies will help in understanding its limitations and effectiveness in real-world settings. The study's findings need to be replicated in larger and more diverse populations to confirm their generalizability. Efforts should be made to integrate voice analysis tools into existing healthcare systems. This includes training healthcare providers and ensuring compatibility with electronic health records.

In summary, the holistic approach provides compelling evidence for the use of machine learning and voice analysis in the early detection and differentiation of cognitive function disorders, offering a promising direction for future research and application in health-care. Our methodology also has potential applications in diagnosing and monitoring other health conditions where vocal changes are indicative, such as Parkinson's disease or even some psychiatric conditions.